\documentclass{emulateapj}

\usepackage{amsmath}

\newcommand{\kms}{km\,s$^{-1}$}

\shorttitle{Post-shock temperature, cosmic-ray pressure and 
cosmic-ray escape.}
\shortauthors{J. Vink  et al.}

\def\pasa{{PASA}}

\begin{document}

\title{
The relation between post-shock temperature, cosmic-ray pressure and 
cosmic-ray escape for non-relativistic shocks
}
\author{Jacco Vink$^1$, Ryo Yamazaki$^2$, Eveline A. Helder$^1$, K. M. Schure$^1$}
\affil{$^1$Astronomical Institute, Utrecht University, P.O. Box 80000, 
3508TA Utrecht, The Netherlands
$^2$ Department of Physics and Mathematics, Aoyama Gakuin University, 5-10-1 Fuchinobe, Sagamihara, Kanagawa, 252-5258, Japan
}

\email{j.vink@astro-uu.nl}

\begin{abstract}
Supernova remnants are thought to be the dominant source of Galactic
cosmic rays. This requires that at least 5\% of the available
energy is transferred to cosmic rays, implying a high
cosmic-ray pressure downstream of supernova remnant shocks.
Recently, it has been shown that the downstream temperature in some remnants
is low compared to the measured shock velocities,
implying that additional pressure support by accelerated particles
is present.
                                   
Here we use a two-fluid thermodynamic approach to derive the relation
between post-shock fractional cosmic-ray pressure and post-shock temperature, 
assuming no additional heating beyond adiabatic heating
in the shock precursor and 
with all non-adiabatic heating occurring at the subshock.
The derived relations show that a high fractional cosmic-ray pressure is
only possible, if a substantial fraction of the incoming energy
flux escapes from the system.  
Recently a shock velocity and a downstream proton temperature
were measured for a shock in the supernova remnant RCW 86. We apply
the two-fluid solutions to these measurements and find that the
the downstream fractional cosmic-ray pressure is at least 50\% with a
cosmic-ray energy flux escape of at least 20\%.

In general, in order to have 5\% of the supernova energy to go into
accelerating cosmic rays, on average the post-shock cosmic-ray pressure
needs to be 30\% for an effective cosmic-ray adiabatic index of
$\gamma_{\rm cr}=4/3$. 
\end{abstract}


\section{Introduction}

Although there have been many theoretical and observational
advances in understanding
cosmic-ray acceleration by supernova remnant (SNR) shocks over the last decade, 
it is still not yet clear whether SNRs are capable of putting
more than 5\% of their energy into cosmic rays. This number is necessary
in order to explain the cosmic-ray energy density in the Galaxy, given
the size of the Galaxy, the supernova rate, and the cosmic-ray escape
time \citep[e.g.][for an early discussion]{ginzburg67}.

The theoretical advances consist, among others,
of a better understanding of
magnetic field amplification by cosmic-ray streaming \citep[e.g.][]{bell04},
and major improvements in the  self-consistent modeling of efficient cosmic-ray 
acceleration \citep[e.g.][]{blasi05,kang09,vladimirov08}.
These simulations show that SNR shocks can transfer up to $\sim 80$\% of
the kinetic energy flux entering the shock to cosmic rays.

On the observational side progress has been made through X-ray observations
which show that many young SNRs are actively accelerating electrons
up to 10-100 TeV, and they show that the magnetic fields must indeed be
amplified \citep[e.g.][]{vink03a,voelk05,bamba05,ballet06}.
Another major source of progress
has been the 
coming of age of TeV gamma-ray astronomy. TeV gamma-ray observations
have shown us that many
young SNRs are TeV sources \citep[e.g.][]{aharonian01,aharonian04,albert07,
acciari10}. This proves that SNRs are capable of
accelerating particles up to at least 100~TeV.
Unfortunately, a correct interpretation of
the emission mechanism, i.e. pion-decay or inverse Compton scattering, 
is necessary for understanding the fraction of energy
contained by accelerated particles \citep[see the review by][]{hinton09}. 

Recent GeV gamma-ray observations with the
Fermi-LAT even show that, whatever the emission mechanism, a bright
young SNR like Cassiopeia A (Cas A) has only transferred 
$\sim 2$\% of its explosion
energy to accelerated particles. The interpretation is not yet clear:
are Cas A and other young SNRs beyond their peak in acceleration power, 
and did most of the highest energy particles escape? 
Or did they not yet reach their full potential as sources of
cosmic rays?
Or are young SNRs not the dominant sources of Galactic cosmic rays?

Recently, another method of measuring cosmic-ray acceleration efficiency
has drawn attention. It consists of measuring the proton temperature
behind SNR shocks that are suspected of being efficient cosmic-ray 
accelerators, in view of their high velocity or multi-wavelength properties.
Without cosmic-ray acceleration there exists a simple relationship
between the post-shock plasma temperature and shock velocity. However,
when a shock accelerates cosmic rays, less energy is available
for plasma heating. 
The proton temperature
can be deduced
from thermal Doppler broadening of H$\alpha$ line emission from
behind fast shocks \citep{heng10}. Note that, unlike the electron
temperature,  the proton temperature is for cosmic abundance
close to or equal to the mean plasma temperature.

This method was used by \citet{helder09} for an X-ray synchrotron
emitting shock in the SNR RCW 86. The X-ray synchrotron emission
indicates that the shock is actively accelerating particles to
$>10$ TeV energies. 
Moreover, RCW 86 is detected by H.E.S.S. as a TeV gamma-ray source
\citep{aharonian08}. Indeed \citet{helder09} found that the
plasma temperature is a factor of at least three lower than expected given
the measured shock velocity. However, this value could not
be directly translated into a fractional cosmic-ray pressure
behind the shock, but was translated into a lower limit of 
fractional cosmic-ray pressure of $\geq 50$\%.
This lower limit was derived using the Rankine-Hugoniot relations
for a two-fluid shock \citep[see also][]{vink08d}.

Here we explore the Rankine-Hugoniot relations for a two-fluid shock
further. We show that for a given fractional cosmic-ray 
pressure $w(\equiv P_{\rm cr}/P_{\rm tot})$, there is a unique cosmic-ray
energy escape flux, $\epsilon_{\rm esc} (= F_{\rm cr}/\frac{1}{2}\rho_0 V_{\rm s}^3)$,
associated with it that only depends on the
overall Mach number of the shock. 

In the next section we present the derivation
of the relation between $w$, $\epsilon_{\rm esc}$ and post-shock plasma
temperature. 
In Sect.~\ref{discussion} we discuss this relation and its limitations
in the context
of two-fluid models, and cosmic-ray acceleration models, and we
use the relations to derive the cosmic-ray pressure content for
the northeastern region of RCW 86 and for the newly measured
plasma temperature of the young Large Magellanic Cloud SNR 0509-67.5
\citep{helder10}.

\section{The relation between escape and pressure of cosmic rays}

Efficient particle acceleration by shock fronts leads to a shock structure
that deviates significantly from a normal ``one fluid'' shock 
\citep[e.g.][]{drury81,berezhko99,blasi05,vladimirov08,drury09,kang09,reville09}:
the particles diffusing ahead (=upstream) of the shock, form a shock precursor
that pre-compresses and slows down the
gas flowing into the shock. 
 The pre-compression caused by the precursor adiabatically
heats the gas. The Mach number at the shock is, therefore,
reduced with respect to the overall Mach number, as the shock velocity
is reduced and the gas pressure upstream of the shock is increased with
respect to a shock without a precursor.
Additional heating of the gas may occur in the precursor 
due to non-adiabatic processes
such as Alfv\'enic heating \citep[e.g.][]{vladimirov08,caprioli08}.
In such a multi-fluid system, the shock that heats the plasma is called
the subshock. In the limit of a one fluid gas the subshock is identical to
the shock.

Here we follow a different approach then \citet{blasi05,vladimirov08,kang09,reville09}
in that
we treat the whole system only thermodynamically, using a two-fluid approach,
with the two components representing the thermal gas and a gas of
accelerated particles (cosmic rays).
For the moment we neglect the possible influence of non-adiabatic heating
in the shock precursor due to interactions between the gas and accelerated
particles. Our approach is reminiscent of the work by \citet{drury81}
\citep[see also][]{achterberg84,voelk84,drury09}
and is an extension of the work presented by \citet{vink08d} and 
\citet{helder09}.

\subsection{The Dependence of Post-shock cosmic-Ray Pressure on 
the Overall and Subshock Compression Ratio}

Our starting point is the basic Equations expressing conservation of
mass, momentum and energy flux. We evaluate these expressions
for three distinct regions: (0) far upstream, where the presence
of accelerated particles can be neglected,
(1) in the shock precursor, just upstream of the subshock, and
(2) behind (downstream of) the shock. Mass flux conservation gives:
\begin{equation}
\rho_0v_0 = \rho_1v_1 = \rho_2 v_2,
\end{equation}
with $v$ the velocity of the gas with respect to the shock. Note that
$v_0=V_{\rm s}$, the shock velocity in the observer's frame of reference.
Momentum flux conservation can be expressed as:
\begin{equation}
P_0 + \rho_0 V_{\rm s}^2 = P_1 + \rho_1 v_1^2 = P_2 + \rho_2 v_2^2,
\label{eq:momentum}
\end{equation}
with $P$ the pressure.

At this point it is convenient to introduce 
the Mach number $M_0$ for the shock structure far upstream,
\begin{equation}
M_0^2 \equiv \frac{1}{\gamma_{\rm g}} \frac{\rho_0V_{\rm s}^2 }{P_0},\label{eq:mach}
\end{equation}
with $\gamma_{\rm g}$ the adiabatic index of the thermal particles.
In addition we introduce the compression ratios across the different
regions 0,1, and 2:
\begin{equation}
\chi_1 \equiv \frac{\rho_1}{\rho_0}, \chi_2 \equiv \frac{\rho_2}{\rho_1},
\chi_{12} \equiv \chi_1\chi_2 = \frac{\rho_2}{\rho_0}.\label{eq:chi}
\end{equation}

We now use the assumption that the thermal pressure in the precursor is only
due to adiabatic heating in the precursor, 
i.e. $P_{1,\rm th}= P_0 \chi_1^{\gamma_{\rm g}}$ \citep[c.f.][]{drury09}, and
that across the
subshock the pressure associated with
the  accelerated particles does not change 
\citep[see][]{drury81,achterberg04}. The cosmic-ray
terms therefore cancel each other
in Equation~(\ref{eq:momentum}), when considering the momentum flux
from region 1 to 2.
In other words the non-thermal particle pressure at the subshock is only
relevant for reducing the Mach number,  but leads otherwise
again to the standard, one fluid, relation for shock compression
with a reduced Mach number given by:
\begin{equation}
M_1^2 = \frac{\rho_1v_1^2}{\gamma_{\rm g}P_1} = \frac{\rho_0V_{\rm s}^2/\chi_1}{\gamma_{\rm g} P_0\chi_1^{\gamma_{\rm g}}} =M_0^2 \chi_1^{-(\gamma_{\rm g}+1)}.\label{eq:M1}
\end{equation}

\begin{figure*}
\centerline{
\includegraphics[angle=-90,width=0.5\textwidth]{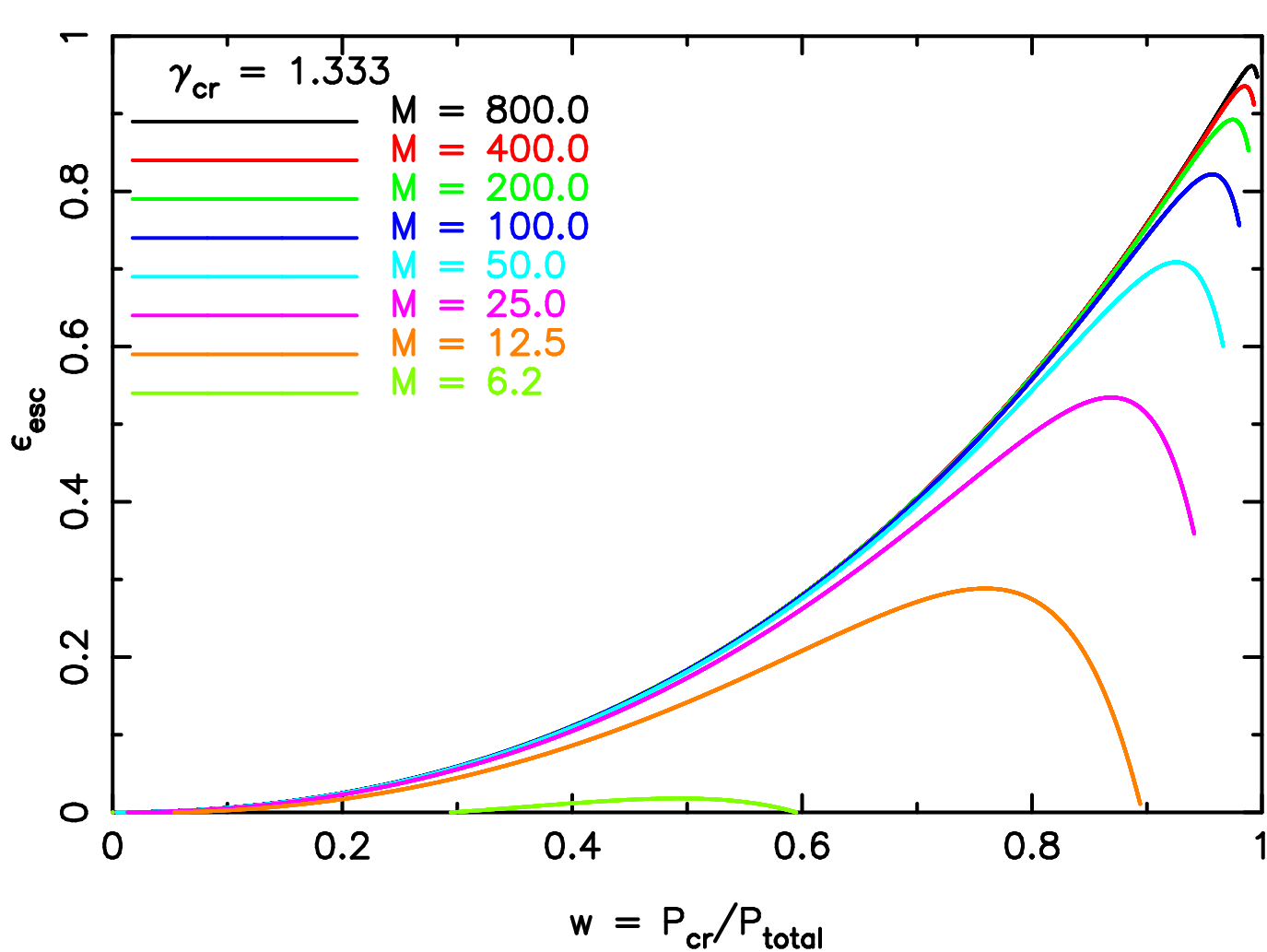}
\includegraphics[angle=-90,width=0.5\textwidth]{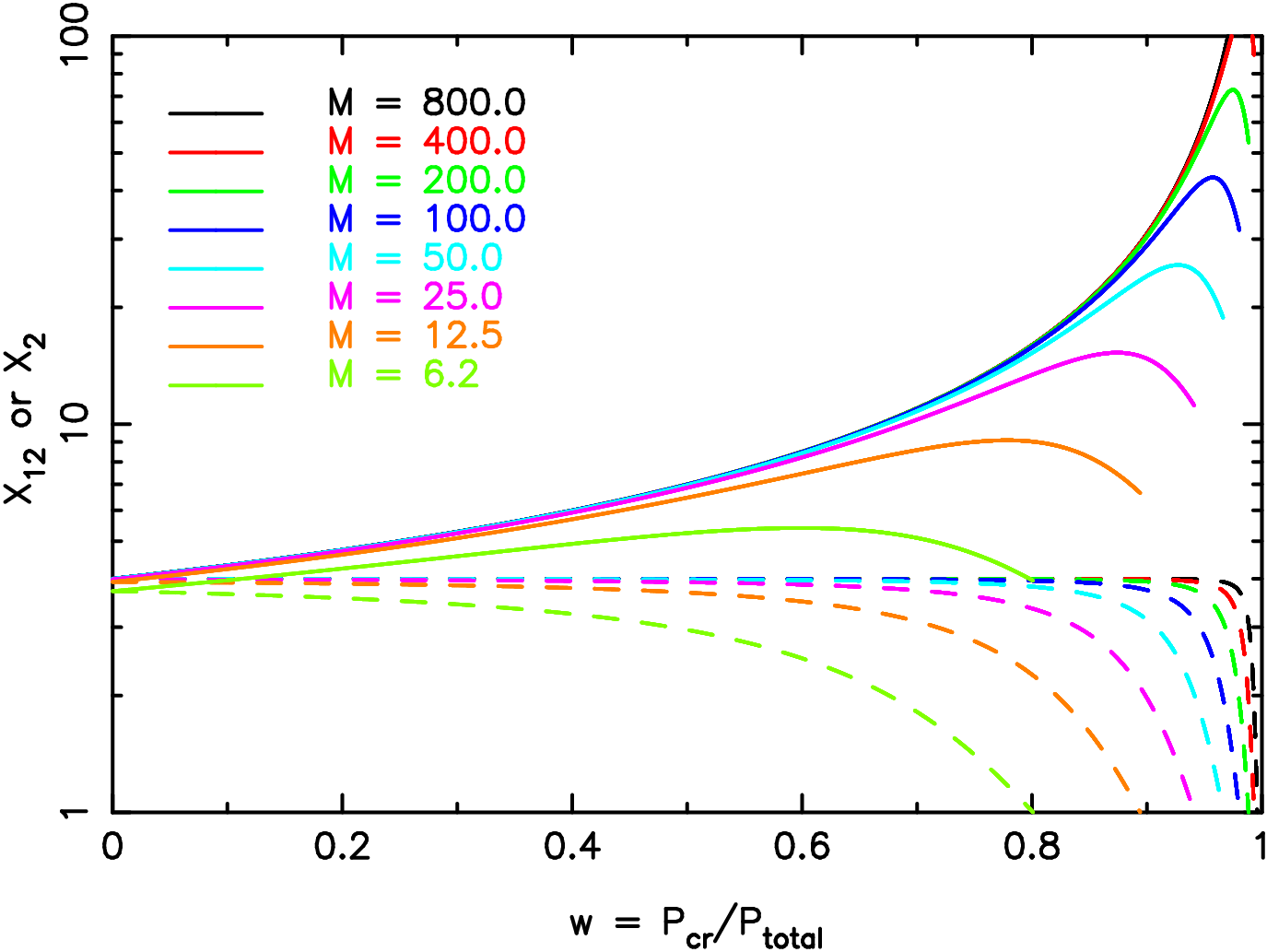}}
\caption{
Left:
The relation between energy flux escape ($\epsilon_{\rm esc}$, 
Equation.~\ref{eq:epsilon}) 
 and downstream fractional cosmic-ray pressure ($w$, Eq.~\ref{eq:w})
for an effective adiabatic index for the cosmic rays of $\gamma_{\rm cr}=4/3$,
and for various values of the
upstream Mach number ($M_0$). Right:
the relation between overall compression ratio ($\chi_{12}$, solid lines)
and subshock compression ratio ($\chi_{2}$, dashed lines).
\label{fig:epsilon}}
\end{figure*}

\begin{figure}
\centerline{
\includegraphics[angle=-90,width=0.5\textwidth]{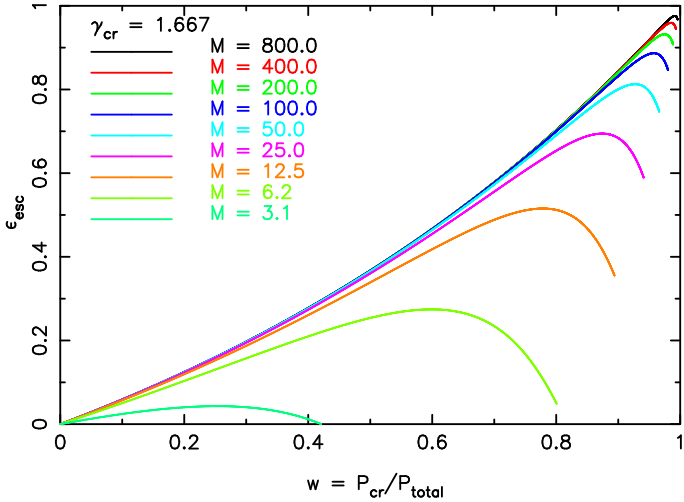}
}
\caption{
Left: The same as the left panel of Figure~\ref{fig:epsilon}(left),
but now for an adiabatic index for the cosmic rays of $\gamma_{\rm cr}=5/3$.
\label{fig:epsilon2}}
\end{figure}

Combining 
Equations~(\ref{eq:momentum}), (\ref{eq:mach}), and (\ref{eq:chi}) one finds for the total 
downstream
pressure $P_2$ and the thermal pressure $P_{th}$:
\begin{equation}
\frac{P_2}{\rho_0V_{\rm s}^2} = \frac{P_{2,cr} + P_{2,th}}{\rho_0V_{\rm s}^2}
= \frac{1}{\gamma_{\rm g} M_0^2} +\Bigl(1 - \frac{1}{\chi_{12}}\Bigr),
\label{eq:P2}
\end{equation}
\begin{equation}
\frac{P_{2,th}}{\rho_1v_1^2}  = (1 - w) \frac{P_2}{\rho_1v_1^2} = 
\frac{\chi_1^{\gamma_{\rm g}+1}}{\gamma_{\rm g}M_0^2} + \Bigl(1 - \frac{1}{\chi_2}\Bigr),
\label{eq:P2th}
\end{equation}
with the subscript $cr$ referring to the cosmic rays.
We have introduced here the symbol $w$ describing the fraction of the
downstream pressure contributed by cosmic rays
\citep[c.f.][]{vink08d,helder09}:
\begin{equation}
w\equiv \frac{P_{2,cr}}{P_{2,th}+P_{2,cr}}.
\end{equation}

Using $\rho_1v_1^2=\rho_0V_{\rm s}^2/\chi_1$ we can now derive
an expression for the fractional cosmic-ray pressure for a given
Mach number $M_0$:
\begin{equation}
w = \frac{(1-\chi_1^{\gamma_{\rm g}}) + \gamma_{\rm g} M_0^2\Bigl(1 - \frac{1}{\chi_1}\Bigr)}
{1 + \gamma_{\rm g} M_0^2 \Bigl(1 -\frac{1}{\chi_{12}}\Bigr)}.\label{eq:w}
\end{equation}
For high Mach number shocks this Equation simplifies to:
\begin{equation}
w = \frac{1 - \frac{1}{\chi_1}}
{1 -\frac{1}{\chi_{12}}}= 
\frac{\chi_{12} - \chi_2}
{\chi_{12} -1}.
\label{eq:w2}
\end{equation}

The compression ratio of the subshock is given by the standard shock relation:
\begin{equation}
\chi_2 = \frac{(\gamma_{\rm g} + 1)M_1^2}{(\gamma_{\rm g}-1)M_1^2 + 2},\label{eq:chi2}
\end{equation}
with $M_1$ being given by Equation~(\ref{eq:M1}).

Equations~(\ref{eq:M1}),(\ref{eq:w}) and (\ref{eq:chi2}) together
show
that
there is a one to one relation between the compression ratio
in the precursor and the downstream fractional cosmic-ray pressure
for a given upstream Mach number $M_0$.

\subsection{The Escaping Energy Flux}
In order to determine the escaping energy flux carried away by
particles diffusing away far upstream, 
we need to use the expression for conservation of energy flux across the shock,
but with a modification in order to express the fact that energy flux
may be lost from  the system
\citep[e.g.][]{berezhko99}:
\begin{equation}
\Bigl\{P_2 + u_2 + \frac{1}{2}\rho_2 v_2^2\Bigr\}v_2 = 
\Bigl\{P_0 + u_0  + (1-\epsilon_{\rm esc})\frac{1}{2}\rho_0V_{\rm s}^2\Bigr\}V_{\rm s},\label{eq:energy}
\end{equation}
with $u=P/(\gamma-1)$ the internal energy, and 
\begin{equation}
\epsilon_{\rm esc} = \frac{F_{\rm cr}}{\frac{1}{2}\rho_0 V_{\rm s}^3},
\end{equation}
the escaping cosmic-ray energy flux, normalized to the total kinetic
energy flux of the shock.
Note that the escaping energy flux can only be taken out of
the kinetic energy flux, as this is the only source of free energy.
If we would have considered radiative losses, then the factor $(1-\epsilon_{\rm esc})$
should have been in front of $P_0$ as well, as the usptream thermal energy
can also be radiated away.

Following \citet{vink08d, helder09} we introduce for convenience 
\begin{equation}
G_0 \equiv \frac{\gamma_{\rm g}}{\gamma_{\rm g} -1},
G_2 \equiv w\frac{\gamma_{\rm cr}}{\gamma_{\rm cr}-1} + (1-w)\frac{\gamma_{\rm g}}{\gamma_{\rm g} -1}.\label{eq:g2}
\end{equation}
Reordering the terms and using Equation~(\ref{eq:P2}) gives:
\begin{equation}
\epsilon_{\rm esc} = 1 + \frac{2G_0}{\gamma_{\rm g} M_0^2} - 
\frac{2G_2}{\gamma_{\rm g} M_0^2 \chi_{12}} - 
\frac{2G_2}{\chi_{12}}\Bigl(1 - \frac{1}{\chi_{12}}\Bigr) - \frac{1}{\chi_{12}^2}.\label{eq:epsilon}
\end{equation}
This Equation completes the thermodynamic relation between the
precursor compression ratio $\chi_1$ and downstream non-thermal pressure and
overall energy flux escape. The resulting relation between $\epsilon_{\rm esc}$
and $w$ can be seen in Figure~\ref{fig:epsilon}.
 For high Mach number shocks the
second and third terms can be omitted. 

In the limit of $\gamma_{\rm cr}=5/3$ and $M_0,M_1 \rightarrow \infty$ one can show that the
relation between $\epsilon_{\rm esc}$ and $w$ is well approximated by:
\begin{equation}
\label{eq:epsw}
\frac{\epsilon_{\rm esc}}{w} = \Bigl( 1 - \frac{1}{\chi_{12}}\Bigr)^2.
\end{equation}
Since  realistically $\gamma_{\rm cr}<5/3$ (see Discussion), 
Equation~(\ref{eq:epsw}) serves as an upper bound on the escape flux.

For low Mach number shocks there is a critical Mach number,
$M_0 \approx 6$
below which cosmic-ray acceleration cannot be thermodynamically
supported for $\gamma_{\rm cr}=4/3$, as can be seen in Figure~\ref{fig:epsilon}.
For $\gamma_{\rm cr}=5/3$ (Figure~\ref{fig:epsilon2}) cosmic-ray acceleration can be supported for lower Mach numbers, but there is still a limit, 
since $\epsilon_{\rm esc} < 0$ for $M_0 \lesssim 2.5$.

To summarize: taking $\chi_1$ as the principal variable, and $M_0$ as
an input parameter, one
can calculate the subshock Mach number $M_1$ (Equation~(\ref{eq:M1})), 
from which 
the subshock compression ratio follows (Equation~(\ref{eq:chi2}), 
and therewith the overall compression ratio $\chi_{12}=\chi_1\chi_2$.
This can then be used to calculate $w$ (Equation~(\ref{eq:w})) and 
$\epsilon_{\rm esc}$ (Equation~(\ref{eq:epsilon})). 
We illustrate this in Figure~\ref{fig:epsilon}.

\subsection{The Maximum Overall Compression Ratio}
Note that although for a given $w$ the escape flux
$\epsilon_{\rm esc}$ can be calculated, the
opposite is not true. The reason is that  $\epsilon_{\rm esc}$ has a maximum.
In fact, mathematically the curves beyond the maximum declines further
then indicated in Figure~\ref{fig:epsilon} (left), but this would correspond
to an unphysical compression factor of  $\chi_2 <1$ at the subshock.
The peak value in the overall compression ratio corresponds to
a maximum in $\epsilon_{\rm esc}$. 
The peak value can be found by differentiating the expression 
for the total compression ratio
(see Equation~(\ref{eq:chi2}))
\begin{equation}
\chi_{12}=
\frac{(\gamma_{\rm g}+1)M_1^2\chi_1}{(\gamma_{\rm g}-1)M_1^2+2} =
\frac{(\gamma_{\rm g}+1)M_0^2\chi_1^{-\gamma_{\rm g}}}{(\gamma_{\rm g}-1)M_0^2\chi_1^{-(\gamma_{\rm g}+1)}+2}
\end{equation}
with respect to $\chi_1$, and setting $d\chi_{12}/d\chi_1=0$. 
This gives:
\begin{equation}
\chi_1 =\Bigl(
\frac{\gamma_{\rm g}-1}{2\gamma_{\rm g}}M_0^2
\Bigr)^{1/(\gamma_{\rm g}+1)}.
\end{equation}
Inserting this in Equation~(\ref{eq:chi2}) using  Equation~(\ref{eq:M1})   shows that $\chi_{12}$ reaches
a maximum for a subshock compression ratio of
\begin{equation}
\chi_{2,max}=\frac{\gamma_{\rm g}}{\gamma_{\rm g}-1}=5/2,
\end{equation}
with the numerical value valid for $\gamma_{\rm g}=5/3$.

\begin{figure*}
\centerline{
\includegraphics[angle=-90,width=0.5\textwidth]{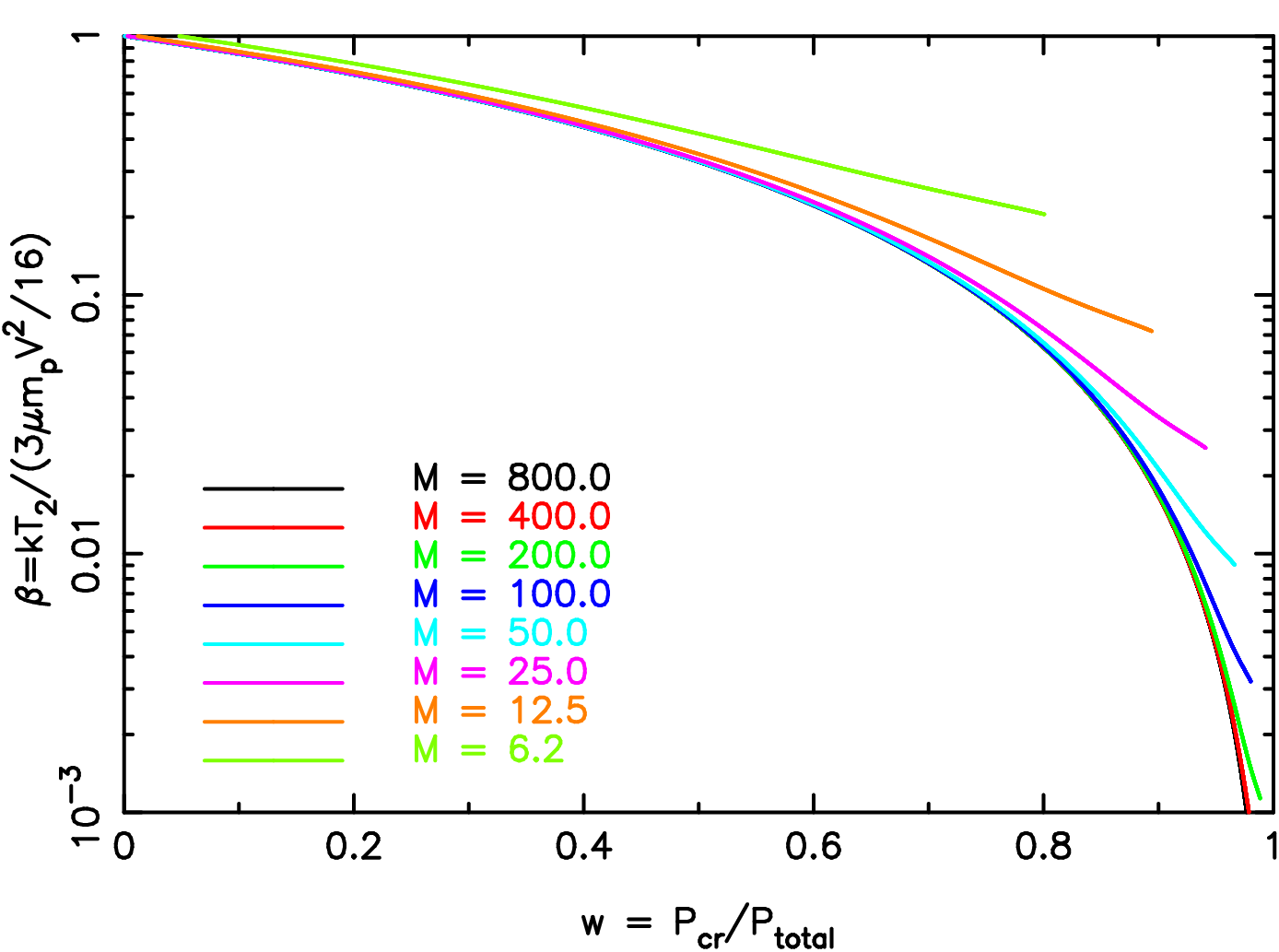}
\includegraphics[angle=-90,width=0.5\textwidth]{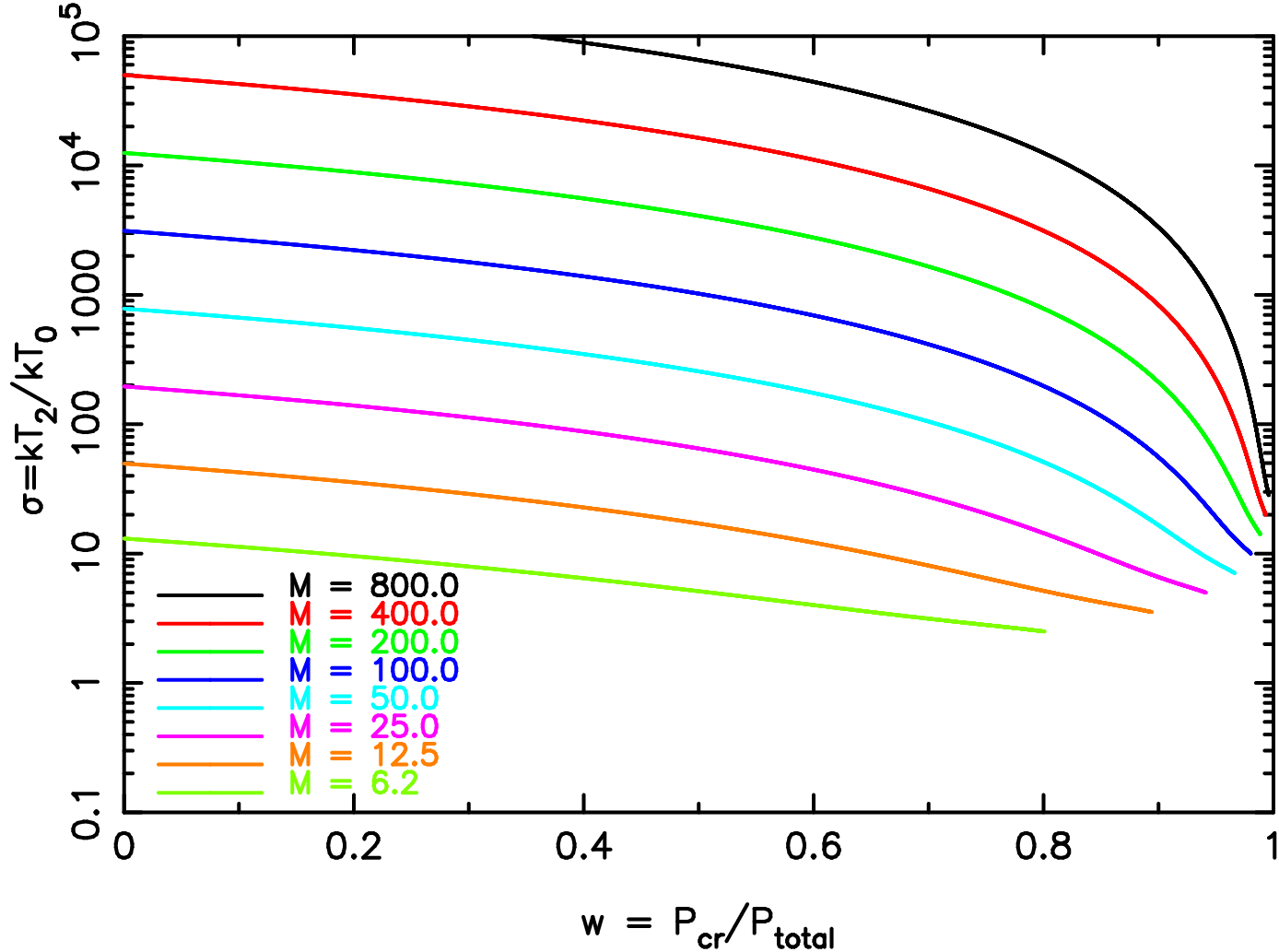}
}
\caption{
Left: The ratio $\beta$ between downstream (post-shock) temperature
in a cosmic-ray dominated shock and a normal, high Mach number gas shock
    as a function of fraction downstream cosmic-ray pressure $w$.
Right: The ratio $\sigma$ between upstream and downstream temperature
as function of $w$.
\label{fig:beta}
}
\end{figure*}

\begin{figure}
 \includegraphics[angle=-90,width=0.5\textwidth]{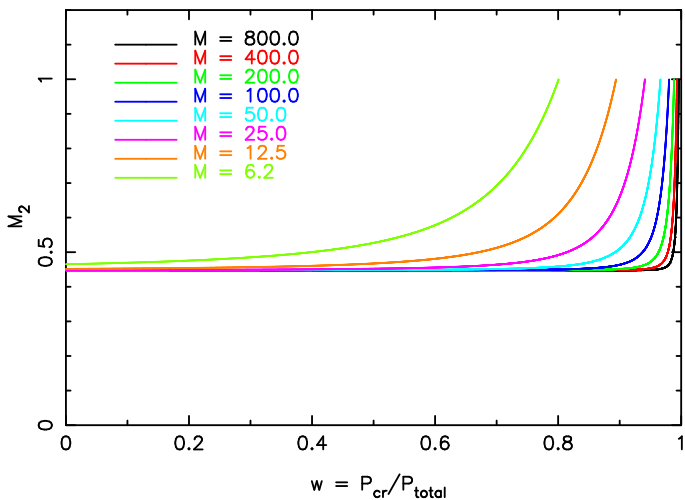}
\caption{
The downstream Mach number $M_2$ as a function of the fractional cosmic
ray pressure $w$.
\label{fig:mach2}
}
\end{figure}

\subsection{The Downstream Temperature}
From Equations~(\ref{eq:P2})  and (\ref{eq:w}) or (\ref{eq:P2th})
it is  possible to find
an expression for the downstream temperature:
\begin{align}
\label{eq:beta}
k_{\rm B}T_2 = &P_{2,th}/n_2 = \\ \nonumber
&(1 - w)\frac{1}{\chi_{12}} \Bigl[ \frac{1}{\gamma_{\rm g} M_0^2} + \Bigl(1 - \frac{1}{\chi_{12}}\Bigr)
\Bigr]
\mu m_{\rm p}V_{\rm s}^2 =\\\nonumber
&
       \frac{1}{\chi_1\chi_{12}} 
\Bigl[ 
\frac{\chi_1^{\gamma_{\rm g}+1}}{\gamma_{\rm g} M_0^2} + \Bigl(1 - \frac{1}{\chi_{2}}\Bigr)
\Bigr]
\mu m_{\rm p}V_{\rm s}^2 ,
\end{align}
with $k_{\rm B}$ the Boltzmann constant and
$\mu$ the mean mass per particle in units of the proton mass $m_{\rm p}$.
This expression should be compared to the temperature expected behind
a strong single-fluid shock:
\begin{equation}
k_{\rm B}T_2 =\frac{1}{\chi_{12}} \Bigl(1 - \frac{1}{\chi_{12}}\Bigr) 
\mu m_{\rm p}V_{\rm s}^2 = \frac{3}{16}\mu m_{\rm p}V_{\rm s}^2,
\end{equation}
which can be obtained from
Equation~(\ref{eq:beta}) by setting $\chi_{12}=4$ (valid for
a strong shock with $\gamma_{\rm g}=5/3$), $w=0$ and $M_0 \rightarrow \infty$.

It is useful to define the ratio, 
\begin{equation}
\beta = \frac{k_{\rm B}T_2}{\frac{3}{16}\mu m_{\rm p}V_{\rm s}^2}
\end{equation} 
between the downstream temperature in the
presence of cosmic-ray acceleration and the expected temperature
for a pure gas shock, as this is a quantity that can be related
to existing measurements \citep[c.f.][]{helder09,helder10}.
The behavior of $\beta$ is shown in Figure~\ref{fig:beta} (left).
Note that Equation~(\ref{eq:beta}), like the expression for $w$
(Equation~\ref{eq:w}), only relies on
momentum conservation. As a result there is a unique relation between
$\beta$ and $w$. Since $w$ does not depend on the adiabatic
index $\gamma_{\rm cr}$ of the accelerated particles, 
$\beta$ does not depend on $\gamma_{\rm cr}$, and, therefore,
$\beta$ does not depend on the energy distribution of the accelerated particles.

In addition, we define the ratio $\sigma =k_{\rm B}T_2/k_{\rm B}T_0$ between downstream
and upstream temperature, as it is a quantity used by \citet{blasi05,drury09}. 
The expression for this quantity is:
\begin{align}
\sigma =  \frac{k_{\rm B}T_2}{k_{\rm B}T_0} = &(1-w)\frac{P_2}{\rho_0V_{\rm s}^2}
\frac{\gamma_{\rm g} M_0^2}{\chi_{12}} = \\ \nonumber
&(1-w) \frac{1}{\chi_{12}}\Bigl[ 
1 + \gamma_{\rm g} M_0^2\Bigl(1 - \frac{1}{\chi_{12}}\Bigr) 
\Bigr] =\\ \nonumber
& \frac{1}{\chi_1\chi_{12}}\Bigl[  \chi_1^{\gamma_{\rm g}+1} +
\gamma_{\rm g} M_0^2 \Bigl(1 - \frac{1}{\chi_2}\Bigr)\Bigr].
\end{align} 
The behavior of $\sigma$ as a function of $w$ is shown in right-hand panel of 
Figure~\ref{fig:beta}.  In addition we show in Figure~\ref{fig:mach2}
the downstream Mach number $M_2$. This shows that for
high $w$ the dowstream Mach number rapidly approaches
$M_2=1$. For $\chi_2=1$, i.e. a continuous shock, we
have $M_2=1$, which leads to an unstable situation, as was
already established by \citet{drury86}.

\subsection{The Potential Influence of Magnetic Field Amplification}
As mentioned in the introduction, there is now observational evidence
for magnetic field amplification by shocks in young SNRs. In our two-fluid
approach we ignore the effects of magnetic fields on the flow parameters,
although we hope to come to this issue in the near future. For now
we offer an assessment of the magnitude of magnetic amplification on
the cosmic-ray dominated shocks. 

We start by noting that
the downstream magnetic field in several young SNRs is consistent with
$(B_2^2/8\pi)/\rho_0 V_{\rm s}^2 \approx 1$~\% \citep{voelk05,vink08d}.
The question now is, what is the perpendicular magnetic field pressure on
the subshock compared to the ram pressure, 
i.e. what is
$(B_{1,\perp}^2/8\pi)/(\rho_1 v_1^2$)?
This can be calculated using the fact that
$B_{1,\perp}= B_{2,\perp}/\chi_2$:
\begin{equation}
\frac{B_{1,\perp}^2/(8\pi)}{\rho_1v_1^2}=
\frac{B_{2,\perp}^2/(8\pi)}{\rho_0V_{\rm s}^2}
\Bigl( \frac{\chi_1}{\chi_2^2}\Bigr)=
\frac{B_{2,\perp}^2/(8\pi)}{\rho_0V_{\rm s}^2}
\Bigl( \frac{\chi_{12}}{\chi_2^3}\Bigr)
\end{equation}
Since $B_{2,\perp}^2<B_2^2$ we can say that observations of young SNRs
indicate that 
\begin{equation}
\frac{B_{1,\perp}^2/(8\pi)}{\rho_1v_1^2} \lesssim 1\%  
\Bigl( \frac{\chi_{12}}{\chi_2^3}\Bigr)
\end{equation}
From Figure~\ref{fig:epsilon} (right) we can see that for a large
range of $w$ the factor $\chi_{12}/\chi_2^3$ is smaller than one. 
Around the maximum of $\chi_{12}$ we have $\chi_2=2.5$. So in order
to have $(B_2^2/8\pi)/\rho_0 V_{\rm s}^2 \gtrsim 10$\% one needs
$\chi_{12} \gtrsim 158$, which is larger than for any of the models depicted in
Figure~\ref{fig:epsilon}.

However, there has been a debate whether $B^2 \propto \rho_0 V_{\rm s}^2$ \citep{voelk05},
or $B^2 \propto \rho_0 V_{\rm s}^3$ \citep{vink08d,tatischeff09}. For the young
SNRs the precise proportionality does not change our conclusion very much, but
in very young SNRs and/or for radio supernovae this may be relevant.
In particular for SN 1993J, which was bright radio supernova, magnetic fields as high
as 50~G have been inferred, and  $B^2/(8\pi) \sim 0.1 P_{\rm cr}$ \citep[e.g.][]{tatischeff09}.  
At such a level the magnetic fields may start to become dynamically important.

Apart from amplifying the magnetic field upstream of the shock, 
cosmic rays in the precursor
may also give rise to non-adiabatic gas heating 
\citep[e.g.][]{vladimirov08,caprioli09}.
The effects of non-adiabatic heating will
be similar to having a lower Mach number shock, 
i.e. non-adiabatic heating in the
precursor limits the maximum overall compression ratio and
limits the maximum possible
fractional cosmic-ray pressure in the downstream region.

\begin{figure*}
\centerline{
\includegraphics[angle=-90,width=0.5\textwidth]{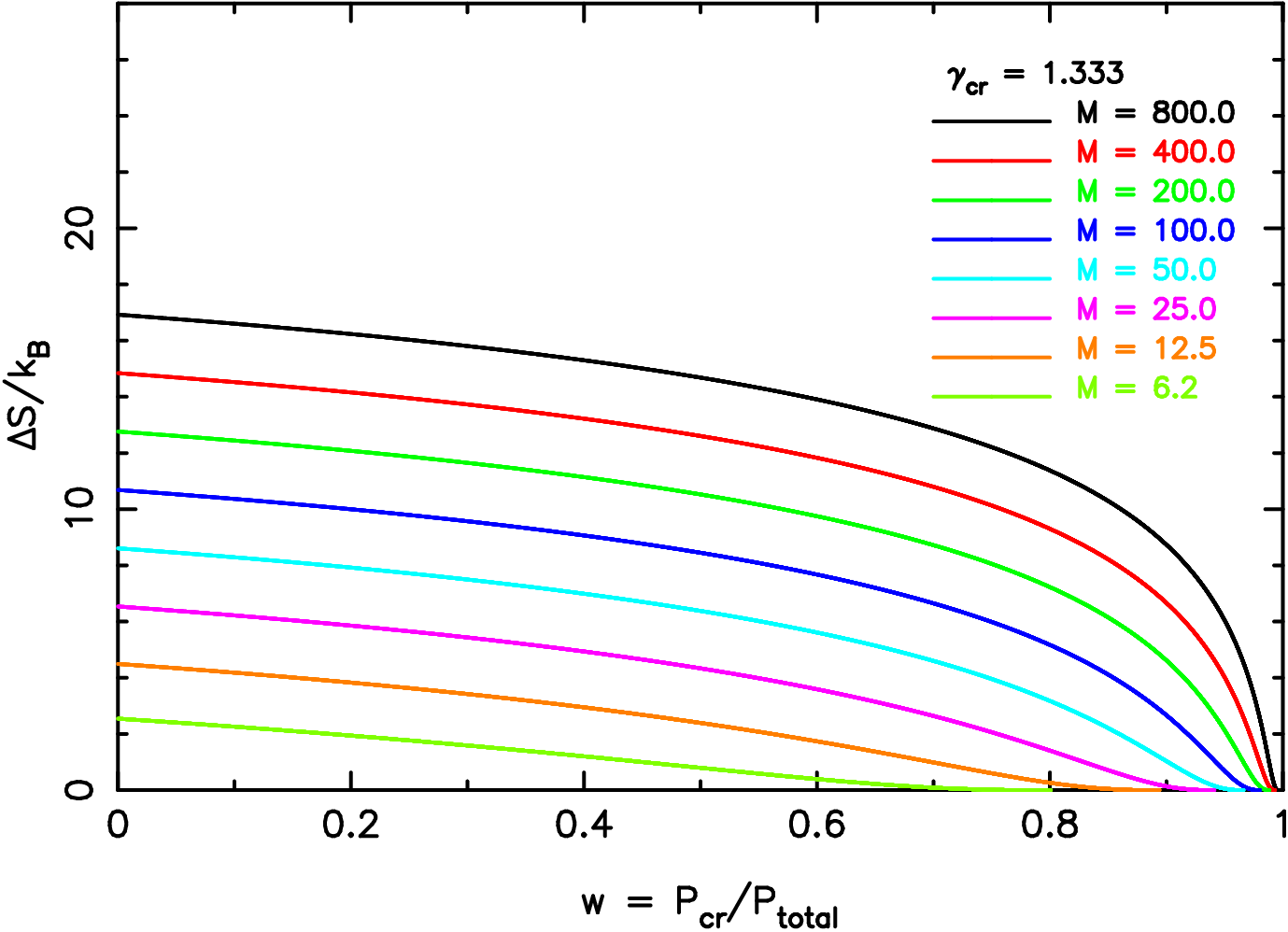}
\includegraphics[angle=-90,width=0.5\textwidth]{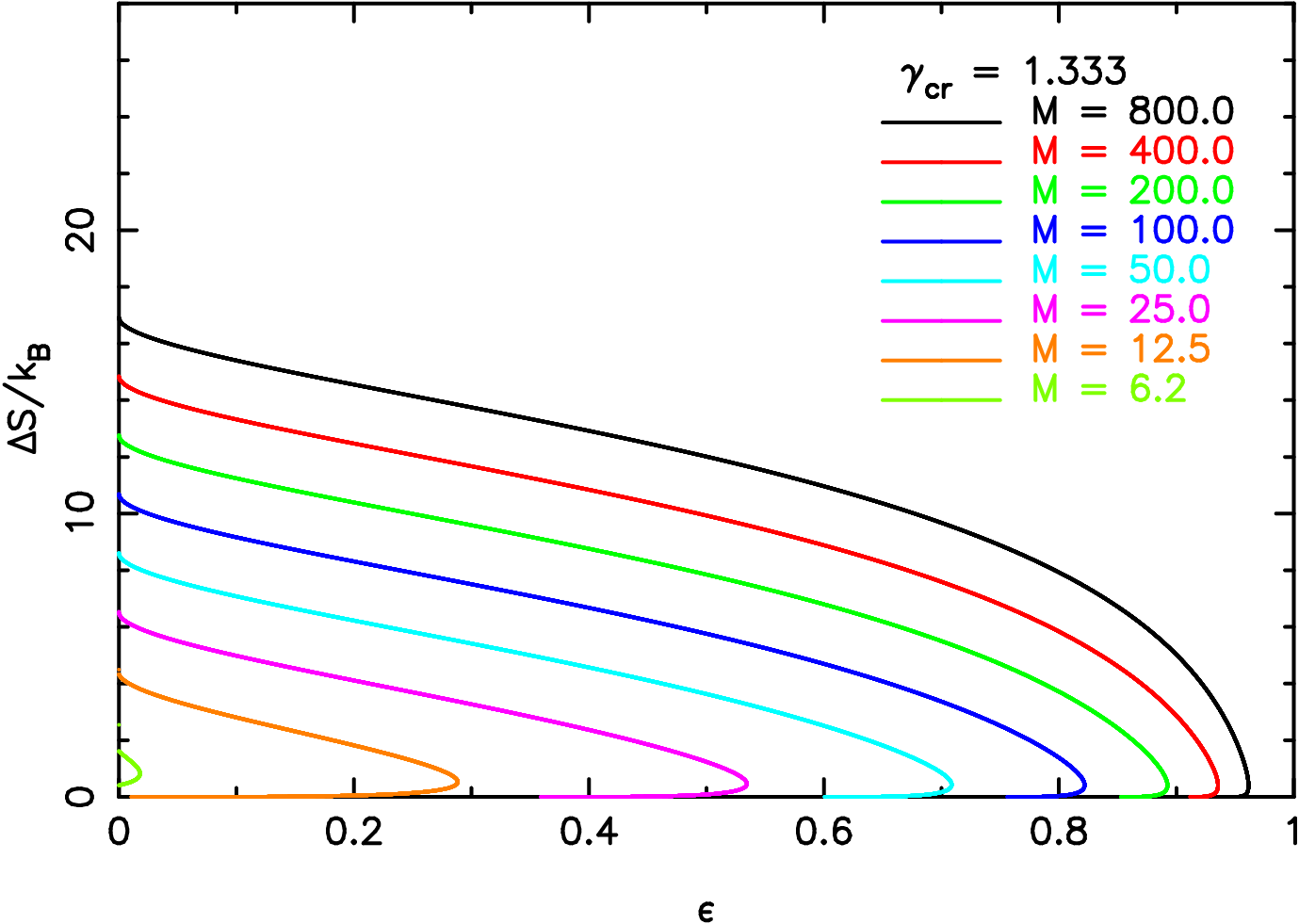}
}
\caption{
The increase in entropy from far upstream to downstream. Left: the entropy
increase as a function of downstream fractional cosmic-ray pressure $w$. 
Right:
idem, but as a function of escaping energy flux $\epsilon_{\rm esc}$.
\label{fig:entropy}
}
\end{figure*}

\subsection{The Entropy Change}
The entropy jump across a shock is given by the relation 
\citep[e.g.][]{zeldovich66}:
\begin{equation}
\Delta S= \frac{3}{2}k_{\rm B} \ln\Bigl(
\frac{P_{2,\rm th}\rho_2^{-\gamma_{\rm g}}}{P_{0,\rm th}\rho_0^{-\gamma_{\rm g}}}\Bigr).\label{eq:entropy}
\end{equation}
This neglects the entropy increase
associated with the accelerated particles. \citet{brown95} calculated
the entropy of non-thermal distributions, which, not surprisingly, 
always have entropy values below a thermal distribution. Moreover, they
normalize the distribution to the number of particles involved. However,
in reality the cosmic-ray component may contribute a dominant
fraction of the pressure, but the number of particles is always much
smaller than  the number of thermal particles. So neglecting the
entropy of the cosmic rays is a good approximation.

Figure~\ref{fig:entropy} (left) shows that, when the fractional cosmic-ray
pressure $w$ increases, the jump in entropy decreases.
The right hand panel is more interesting as it shows that for
certain values of the escaping energy flux $\epsilon_{\rm esc}$ there are two 
possibilities for the jump in entropy. The maximum value of $\epsilon_{\rm esc}$
as a function of entropy change $\Delta S$ again occurs for
the value for which $\chi_{12}$ has a maximum. 
One can speculate that whenever there are  possible solutions to
obtain a certain value for $\epsilon_{\rm esc}$ the nature chooses the one that
offers the highest entropy jump. On the other hand the escape has to
be facilitated by the details of the acceleration process and particle
spectrum, so one should not discount the low entropy branch too easily.

\section{Discussion}
\label{discussion}
The solutions that we presented here for the two-fluid approach to
shock with a cosmic-ray component give us a  handle on estimating the
effects of cosmic-ray acceleration on the post-shock plasma temperature, and
in addition allow us to estimate the cosmic-ray escape. It may not be too surprising
that there is a thermodynamic relation between fractional cosmic-ray pressure $w$
and escape. The reason is that in order to have a shock jump one needs either viscous
shock, or one needs to lose energy from the system. The problem with a cosmic-ray
component is that the accelerated particles do not cause a jump in entropy. Therefore,
the increased overall shock compression has to be facilitated by energy escape.

This situation will change once additional heating in the precursor caused by
cosmic-ray heating is introduced. This heating is presumably the result of
dissipation of large-amplitude 
magnetohydrodynamic waves in the cosmic-ray precursor (Alfv\'enic heating);
these waves 
are the result of cosmic-ray streaming \citep{vladimirov08,caprioli09}.
The most important effect of heating in the precursor will be that
the Mach number at the subshock $M_1$ will be decreased. So we
expect that including precursor heating will resemble the solutions for
lower Mach number shocks.
However, the physical details of cosmic-ray heating are not well
known, and an in depth study of cosmic-ray heating is beyond the scope of 
this paper.
Note that in principal additional heating can be incorporated 
in the Equations,
by parameterizing it in relation to the upstream Mach number and $w$.

Another limitation of the two-fluid approach is that it assumes a steady state 
situation and a plane parallel geometry.
In a non-steady state situation the shock relations may be influenced
by shock acceleration in the past \citep{drury95}. Energy flux conservation
(Equation~\ref{eq:energy})
may therefore be violated.
Moreover, the highest energy particles may take such a long time between two 
consecutive
shock crossings that the shock velocity has appreciably slowed down. 
An early discussion of time dependent effects, as well as the influence
of spherical expansion, can be found in \citet{drury95} 
\citep[see also][]{berezhko94}. Recently 
non-steady
situations were investigated by \citet{kang09}, using a kinetic approach,
and \citet{schure10}, using Monte-Carlo simulations of
test particles coupled to hydrodynamic simulations of 
spherically expanding SNRs.

In non-steady situations high compression ratios can be reached, even
in the absence of escape of cosmic rays from the system, 
because in the precursor there is a larger energy flux associated with
particles diffusing in the  upstream direction than in the downstream direction.
In a steady state situation, this asymmetry
is caused by escape of the highest energy particles. 
In a non-steady state situation, in which the cosmic-ray particle
population is building up \citep{kang09}, this asymmetry
is caused by the fact that the flow in the upstream direction is from a more energetic particle
population that left the sub-shock more recently. It would be interesting to
investigate whether one can model non-steady state shock acceleration
in the two fluid approach by broadening the
definition of  $\epsilon_{\rm esc}$ so that it includes this flux asymmetry.

Note that pressure equilibrium (Equation~\ref{eq:momentum})  is probably
a good approximation even in non-steady state situations.  
As a result, the relation
between $k_{\rm B}T$ and $w$ is expected to be valid even in non-steady
state situations.

\begin{figure*}
\centerline{
\includegraphics[angle=-90,width=0.5\textwidth]{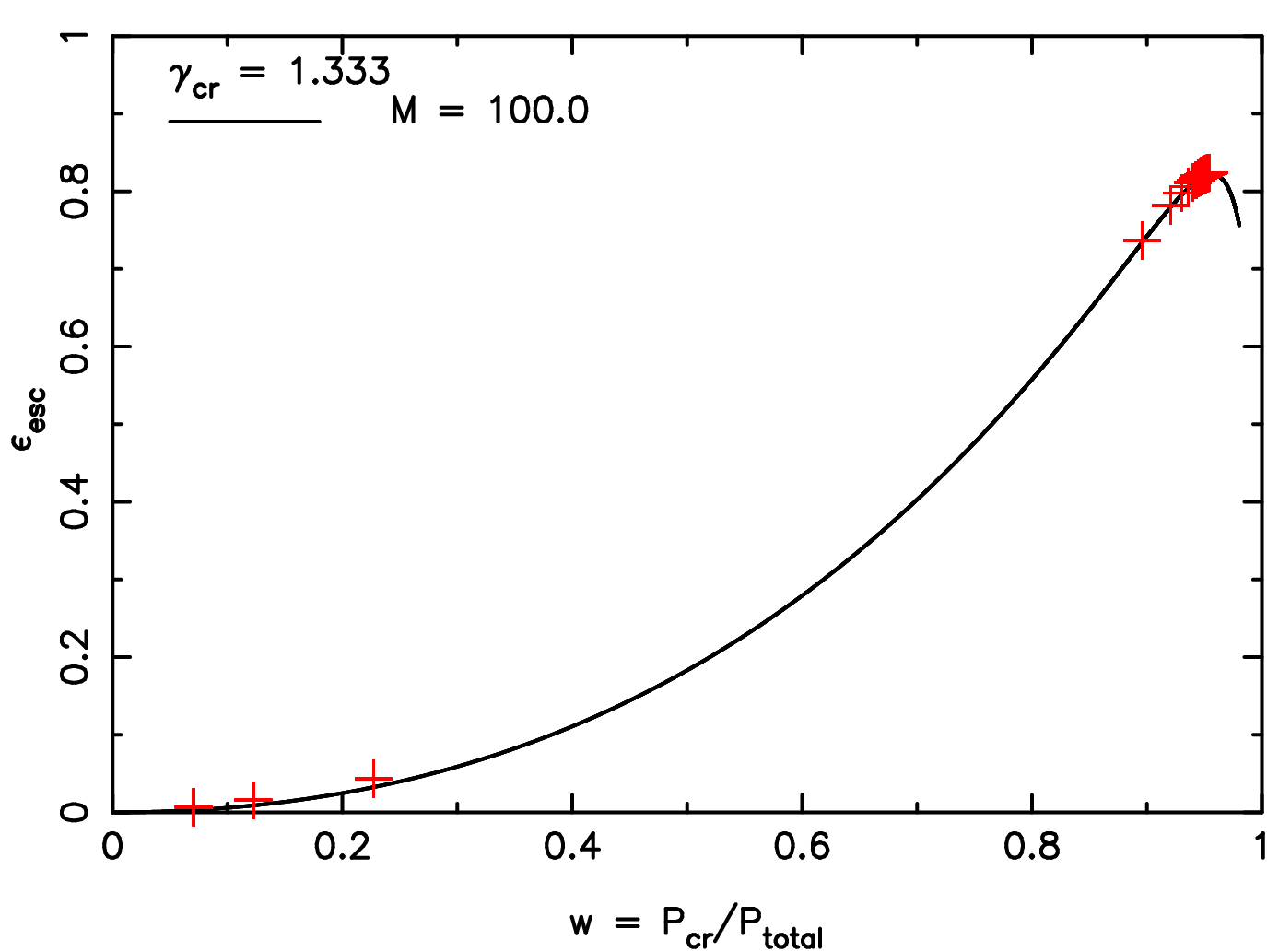}
\includegraphics[angle=-90,width=0.5\textwidth]{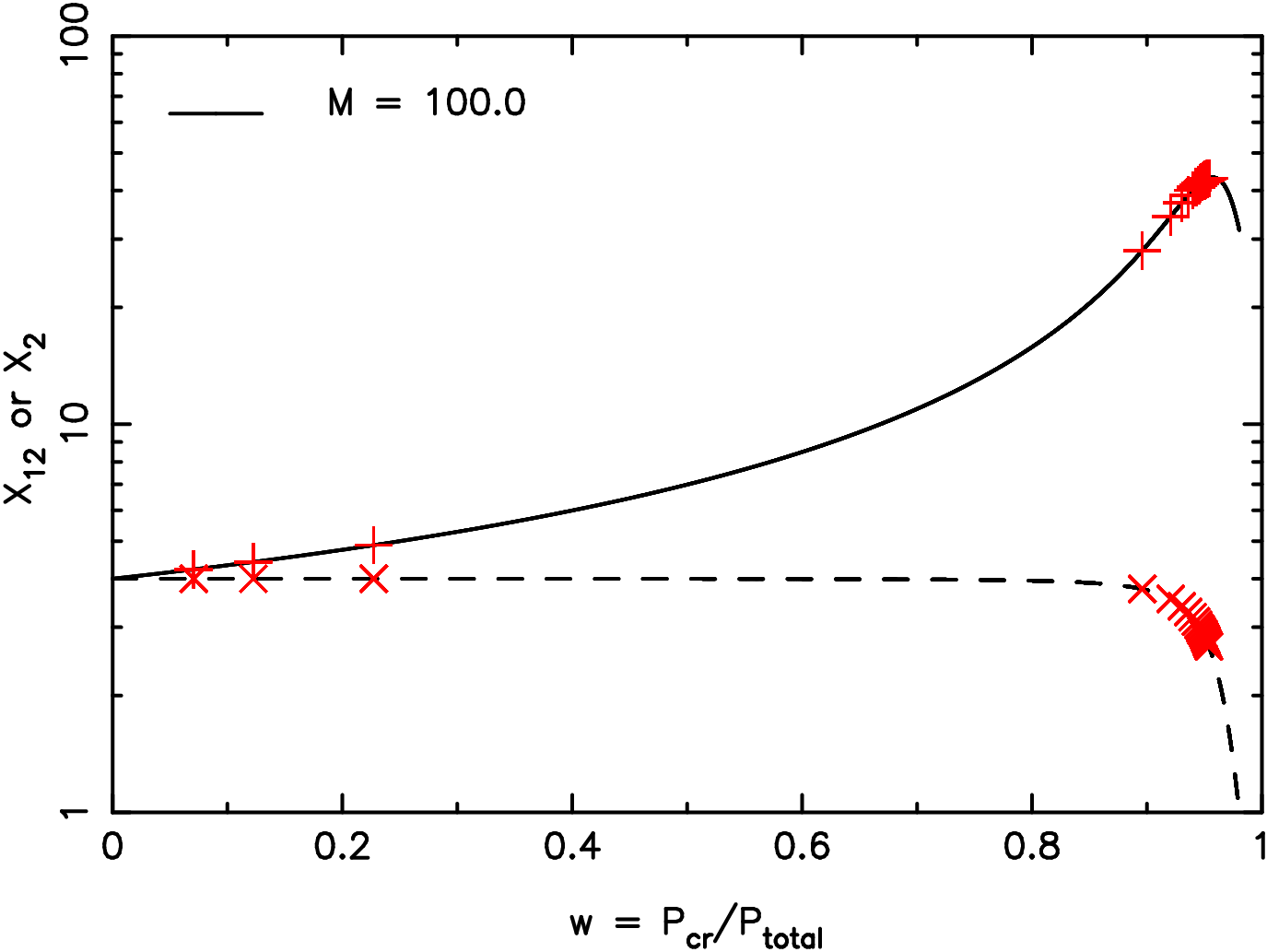}}
\caption{\label{fig:blasi}
Comparison of the two-fluid model with $M_0=100$ with the kinetic
model of \citet{blasi05} (crosses). The kinetic model was
used with $p_{max}=10^5 mc$, $\xi = 3.5$, and for 20 logarithmically 
spaced  velocities in the range of $10-10^5$~\kms.
Left: cosmic-ray escape version
post-shock partial cosmic-ray pressure. Right: overall and subshock compression ratio.
}
\end{figure*}

\subsection{Thermodynamics versus Acceleration Properties}
In the present model the quantities $\chi_2$, $\chi_{12}$, $w$, $\beta$, 
and $\sigma $ are independent on $\gamma_{\rm cr}$, and 
encompass the solutions for particle acceleration based
on nonlinear kinetic models \citep{blasi05,vladimirov08,kang09,reville09},
in which the same Equations of the mass flux and the momentum flux
conservation are adopted. Typically the kinetic models depend on two
free parameters: the injection efficiency and the maximum momentum,
beyond which particles will escape from the system.

As can be seen in Fig.~\ref{fig:blasi} our formalism agrees with the
results of \citet{blasi05}, but the solutions by \citet{blasi05} all
cluster around $w\approx 0.9$ and $\chi_{12}$ around the maximum
possible compression factor. The two-fluid approach, however, allows
for a broader range of solutions.

In the two-fluid model $\epsilon_{\rm esc}$ depends on $\gamma_{\rm cr}$,
which is essentially a free parameter. 
In the kinetic models $\epsilon_{\rm esc}$  is determined self-consistently 
by the overall hydrodynamic
structure of the shock, including the dynamical effects of the
accelerated particles. The energy distribution of the accelerated
particles determines  then $\gamma_{\rm cr}=1 + P_{\rm cr}/u_{\rm cr}$, where
$P_{\rm cr}$ is the cosmic-ray pressure and $u_{\rm cr}$ the cosmic-ray energy density.
Hence the kinematic models predict that
the $\epsilon_{\rm esc}-w$ diagram  should deviate from ours with 
$\gamma_{\rm cr}=4/3$,
and should lie between our curves of $\gamma_{\rm cr}=4/3$ and $5/3$.
In practice, however, the cosmic-ray adiabatic index as obtained by
the kinetic models is very close to $\gamma_{\rm cr}=4/3$ for $p_{max}= 10^5mc$,
as can be seen in  Fig.~\ref{fig:blasi}. For $p_{max}= 10^2mc$
we found that a modest increase to $\gamma_{\rm cr}=1.4$
resulted in a good match
between the two-fluid solutions and the kinetic model of \citet{blasi05}.

Interestingly, the two-fluid solutions presented here indicate that a higher
energy flux escape is necessary for  $\gamma_{\rm cr}>4/3$, but
from the point of view of the spectral energy distribution the energy escape
flux is more difficult to achieve for spectra with spectral energy indices
$\Gamma > 2$ ($N(E)\propto E^{-\Gamma})$, corresponding to $\gamma_{\rm cr}>4/3$.
The reason is that for $\Gamma > 2$ most of the cosmic-ray energy is contained
by mildly relativistic particles, whereas most of the escaping particles will 
be near the maximum of the energy distribution ($p_{max}$ in the
Blasi model). Hence, the escaping particles carry
away only a small fraction of the internal energy. For $\Gamma < 2$ most
energy is indeed concentrated around $p_{max}$ and the necessary
escape flux is easily generated. 

So from a thermodynamic point of view, a high fractional cosmic-ray pressure $w$,
requires large values of $\epsilon_{\rm esc}$, which in turn is best achieved with
$\gamma_{\rm cr} \approx 4/3$.

\subsection{Application to Observations}
One of the motivations for this work has been the 
measurements of the downstream temperature
of SNRs with reasonably well known velocities \citep{helder09,helder10}. 
These temperatures were measured using the broad component of
the H$\alpha$ line. This component is due to charge exchange between
neutral hydrogen atoms entering the shock and the downstream population
of shock heated protons. Hence, the width of the line is caused by
thermal Doppler broadening and reflects the downstream proton temperature.
There is some ambiguity as to how to relate the proton temperature to the overall
downstream plasma temperature, as the different plasma constituents 
(electrons, protons , helium, other ions) may not be in thermal equilibrium. However,
the proton temperature is always expected to be within a factor of 2 of the mean plasma
temperature.

Using Equation~(\ref{eq:beta}) (see also Fig.~\ref{fig:beta}) one can easily
estimate from a measured temperature and shock velocity what
the downstream fractional cosmic-ray pressure is and what the
required escape flux is. The only ambiguity that is left for estimating 
$\epsilon_{\rm esc}$ 
is what the effective cosmic-ray adiabatic index is. However, $\gamma_{\rm cr}$
is not important for determining $w$.
For the SNRs under consideration
the shock velocities are well in excess of 1000~\kms. 
For a typical sound speed in the interstellar medium of 10~\kms, 
we have  $M_0 > 100$. This means that  the high Mach number approximation is valid as long as $w \lesssim 0.9$. 
This appears to be the case for the SNRs considered below.

For the northeastern region of the TeV gamma-ray emitting remnant
RCW 86 \citet{helder09} measured
a downstream temperature of $k_{\rm B}T_{\rm p}= 2.3\pm 0.3$~keV for a measured
shock velocity of $V_{\rm s}=6000 \pm 2800$~\kms. Nominally this corresponds
to $\beta = 0.055$ (see Equation~(\ref{eq:beta})), but given the systematic
uncertainties and the ambiguity due to
non-equilibration of temperatures $\beta$ could be as high  $\beta = 0.31$.
Using now Equations~(\ref{eq:w}) and (\ref{eq:epsilon}) in the limit for
high Mach numbers the measured values of 
$\beta$ correspond to downstream fractional cosmic-ray pressures
and escape fractions of
 $w=0.81, \epsilon_{\rm esc}=0.59$ and $w=0.51, \epsilon_{\rm esc}=0.20$, for 
 $\gamma_{\rm cr}=4/3$ and
 $\beta=0.055$ and $\beta =0.31$ respectively. 
For $\gamma_{\rm cr}=5/3$ this is
$\epsilon_{\rm esc}=0.72$ and $\epsilon_{\rm esc}=0.38$, 
respectively, with $w$ unchanged.

For the young Large Magellanic Cloud remnant 0509-67.5  \citet{helder10}
determined the post-shock temperatures in two regions. The most constraining
measurement was for the southwestern region for which 
$k_{\rm B}T_{\rm p}= 28.7$~keV,
for a conservative velocity estimate of $V_{\rm s}=5000$~\kms, corresponding to
$\beta=0.58$. This translates into a downstream fractional cosmic-ray 
pressure of $w=0.29,\epsilon_{\rm esc}=0.06$ for  $\gamma_{\rm cr}=4/3$ or 
$\epsilon_{\rm esc}=0.19$ for  $\gamma_{\rm cr}=5/3$ .

Interestingly, these new estimates of $w$
based on  Equations~(\ref{eq:w}) and (\ref{eq:epsilon})
are not far from the lower limits given by \citet{helder09,helder10}.
The reason is that  the less constraining relations used by
\citet{helder09,helder10} allow in principle for a higher
fractional cosmic-ray
pressure by reducing the cosmic-ray escape flux. 
The relations derived here do not allow for this possibility.
Moreover the relation
that was used to determine the lower limit by \citet{helder09,helder10}
is a limiting 
case of the shock relations presented here (see Equation~(\ref{eq:epsw})).

\subsection{Do Very Efficiently Accelerating Shocks Exist?}
This brings us back to one of the principle issues currently discussed
in cosmic-ray physics, namely do very efficiently accelerating shock exist?
The post-shock temperatures measured by \citet{helder09,helder10}
indicate high values for the cosmic-ray pressure, but not as high
as usually found by particle acceleration models, where the efficiency
is often found to be close to 90\%, with total compression factors
as high as $\chi_{12}=70$ \citep[e.g.][]{blasi05}. 
It has been argued by \citet{drury09} 
that such extreme conditions may exist in the prominent TeV gamma-ray
source SNR RX J1713.7-3946 \citep{aharonian04}.
Apart from high gamma-ray luminosity this idea is based
on the lack of detectable thermal X-ray emission. According to
 \citet{drury09}  this could be due to a very low post-shock plasma
temperature, possibly as low as low as six times the upstream.
Alternatively,  RX J1713.7-3946 may evolve in a low density medium,
resulting in a low thermal luminosity, as the thermal luminosity scales
with the square of the density. The Equations derived here allow for very
low downstream temperatures. However, the claim by \citet{drury09} that the
downstream temperature may only be six times the upstream temperature seems
unrealistic, as  Equation~(\ref{eq:w}) and
(\ref{eq:epsilon}) show that this is only true for unrealistically low
Mach numbers.
For more realistic Mach numbers for young SNRs, say $M>75$, 
the downstream temperature is at least 80 times the upstream
temperature.

It could be that the young SNRs that we observe are already past their prime
as cosmic-ray accelerators, although the cosmic-ray escape in RCW 86 appears
still high enough to explain the cosmic-ray energy production of SNRs.
Specifically it has been argued that in the very early SNR phase/supernova phase
cosmic-ray acceleration may be very efficient \citep[e.g.][]{bell01,vink08d}, see for example
the case of SN 1993J \citep{tatischeff09}.
Note that a steady escape of cosmic rays during the evolution of a SNR
is favored over a more sudden release of cosmic-rays at the end of the SNR evolution,
as in the latter case adiabatic losses have lowered the energy of the cosmic-rays
\citep{voelk84}.
If one wants to explain the cosmic-ray energy density in the Galaxy
by cosmic-ray escaping from young SNR shocks (i.e. $\epsilon_{\rm esc} = 5$\%, see 
introduction) then on average over the lifetime of the SNR
a fractional cosmic-ray pressure is necessary of at least
$w \approx 30$\% for $\gamma_{\rm cr}=4/3$.

\section{Summary and conclusions}
We have shown that for a two-fluid steady state shock consisting of a thermal gas plus a cosmic-ray component
there exist a unique relation between the downstream fraction
of pressure provide by cosmic rays ($w$) on the one hand, and
the energy escape flux normalized to the incoming
free energy flux ($\epsilon_{\rm esc}$) on the other hand. This relation 
depends on the assumed effective adiabatic index of the cosmic rays and on
the overall Mach number. For Mach numbers $M_0 \lesssim 6$ no
cosmic-ray pressure is allowed. This could be of interest for
particle acceleration in clusters of galaxies,
since internal cluster shocks tend to be low Mach
number shocks, whereas the accretion shocks or cluster formation shock are 
high Mach number shocks.

This relation can be used
to determine $w$ and $\epsilon_{\rm esc}$ based on  measurements of
the downstream temperature of the plasma for a known shock velocity.
Using this relation we show that the {\em lower limits} on $w$ 
inferred for shock regions in the supernova remnants RCW 86 and SNR 0509-675
by \citet{helder09,helder10} are in fact very close to the actual value
for $w$, but we can now determine also $\epsilon_{\rm esc}$, which is at least
$\epsilon_{\rm esc}= 0.2$ for the northeastern region of RCW 86, but only 
0.067 for SNR 0509-675 (but it may be 0.19 in the unlikely case that 
the cosmic rays
have an effective adiabtaic index of $\gamma_{\rm cr}=5/3$).

In order to explain the cosmic-ray energy density in the Galaxy by
cosmic-rays escaping directly from SNR shocks, instead of released
at the end of the SNR life, one needs an escape flux of 
$\epsilon_{\rm esc}\approx 5$\%,
which requires that on average 30\% of the downstream SNR pressure should
be supplied by cosmic rays.

\acknowledgements
JV is supported by a Vidi grant from the Netherlands Science Foundation (NWO).
RY is supported by grant-in-aid from the Ministry of
Education, Culture, Sports, Science, and Technology (MEXT) of 
Japan, No. 19047004, No. 21740184, No. 21540259.

\end{document}